\documentclass[aps, reprint, superscriptaddress, longbibliography, prr]{revtex4-2}

\usepackage{graphicx}
\graphicspath{{images/}}

\usepackage{bm}
\usepackage{dsfont}
\usepackage[dvipsnames]{xcolor}
\usepackage{pstricks}
\usepackage{verbatim}
\usepackage{units}
\usepackage{natbib}
\usepackage{mathtools}
\usepackage{multirow}
\usepackage{enumitem}
\usepackage{mathrsfs}
\usepackage{leftidx}
\usepackage{xspace}
\usepackage{amsmath}
\usepackage[normalem]{ulem} 
\usepackage{soul}
\usepackage{physics}
\usepackage{tabularx}
\usepackage{upgreek}

\usepackage[normalem]{ulem} 
\newcolumntype{Y}{>{\centering\arraybackslash}X}

\usepackage{hyperref}
\hypersetup{colorlinks=true,linktoc=all,linkcolor=blue,breaklinks=true,citecolor=blue,urlcolor=blue}

\usepackage[english]{babel}
\newcommand{\m}{\text{m}}

\newcommand{\In}{\text{in}}
\newcommand{\Out}{\text{out}}

\newcommand{\zpm}{z_\text{ZPM}}
\newcommand{\weff}{\omegam^\text{eff}}
\newcommand{\Gb}{\Gamma_\text{tot}}
\newcommand{\G}{\Gamma}
\newcommand{\g}{\gamma}
\newcommand{\gm}{g_\m}
\newcommand{\Vs}{V_\mathrm{s}}
\newcommand{\gL}{\g_\mathrm{L}}
\newcommand{\gR}{\g_\mathrm{R}}
\newcommand{\GL}{\G_\mathrm{L}}
\newcommand{\GR}{\G_\mathrm{R}}
\newcommand{\muL}{\upmu_{\mathrm{L}}}
\newcommand{\muR}{\upmu_{\mathrm{R}}}
\newcommand{\LR}{\mathrm{L/R}}
\newcommand{\omegam}{\omega_\mathrm{m}}
\newcommand{\etal}{\emph{et al.}}

\newcommand{\mean}[1]{\langle #1 \rangle}

\addto\captionsenglish{}

\makeatletter
\renewcommand\paragraph{\@startsection{paragraph}{4}{\z@}%
  {3.25ex \@plus1ex \@minus.2ex}%
  {-0em}%
  {\normalfont\normalsize\itshape\indent}}
\begin{document}

\title{Ultrastrong coupling between electron tunneling and mechanical motion}
\author{Florian Vigneau} 
\thanks{These authors contributed equally to this work}
\affiliation{Department of Materials, University of Oxford, Oxford OX1 3PH, United Kingdom}

\author{Juliette Monsel}
\thanks{These authors contributed equally to this work}
\affiliation{Department of Microtechnology and Nanoscience (MC2), Chalmers University of Technology, S-412 96 G\"oteborg, Sweden\looseness=-1}

\author{Jorge Tabanera}
\affiliation{Department of Structure of Matter, Thermal Physics and Electrodynamics and GISC, Universidad Complutense de Madrid, Pl. de las Ciencias 1. 28040 Madrid, Spain}

\author{Kushagra Aggarwal} 
\affiliation{Department of Materials, University of Oxford, Oxford OX1 3PH, United Kingdom}

\author{L\'ea Bresque}
\affiliation{Univ. Grenoble Alpes, CNRS, Grenoble INP, Institut N\'eel, 38000 Grenoble, France}

\author{Federico Fedele}
\affiliation{Department of Materials, University of Oxford, Oxford OX1 3PH, United Kingdom}

\author{Federico Cerisola}
\affiliation{Department of Materials, University of Oxford, Oxford OX1 3PH, United Kingdom}
\affiliation{Physics and Astronomy, University of Exeter, Exeter EX4 4QL, United Kingdom}

\author{G.A.D. Briggs}
\affiliation{Department of Materials, University of Oxford, Oxford OX1 3PH, United Kingdom}

\author{Janet Anders}
\affiliation{Physics and Astronomy, University of Exeter, Exeter EX4 4QL, United Kingdom}
\affiliation{Institut f\"ur Physik, Potsdam University, 14476 Potsdam, Germany}

\author{Juan M.R. Parrondo}
\affiliation{Department of Structure of Matter, Thermal Physics and Electrodynamics and GISC, Universidad Complutense de Madrid, Pl. de las Ciencias 1. 28040 Madrid, Spain}

\author{Alexia Auff\`eves}
\affiliation{Univ. Grenoble Alpes, CNRS, Grenoble INP, Institut N\'eel, 38000 Grenoble, France}

\author{Natalia Ares}
\affiliation{Department of Materials, University of Oxford, Oxford OX1 3PH, United Kingdom}

\date{\today}

\begin{abstract}
The ultrastrong coupling of single-electron tunneling and nanomechanical motion opens exciting opportunities to explore fundamental questions and develop new platforms for quantum technologies. We have measured and modeled this electromechanical coupling in a fully-suspended carbon nanotube device and report a ratio of $\gm/\omegam = 2.72 \pm 0.14$, where $\gm/2\pi = 0.80\pm 0.04$~GHz is the coupling strength and $\omegam/2\pi=294.5$~MHz is the mechanical resonance frequency. This is  well within the ultrastrong coupling regime and the highest among all other electromechanical platforms. We show that, although this regime was present in similar fully-suspended carbon nanotube devices, it went unnoticed. Even higher ratios could be achieved with improvement on device design.
\end{abstract}
\maketitle

\setlength{\tabcolsep}{4pt}
\renewcommand{\arraystretch}{1.3}

\section{Introduction}

Ultrastrong coupling between a quantum system and a nanomechanical resonator is reached when the ratio between the coupling strength $\gm$ and the mechanical resonance frequency $\omegam/2\pi$ is greater than one.
In the dispersive regime, such high coupling opens a wide range of possibilities for the development of promising applications in quantum information processing \cite{LaHaye2009}, high precision sensors \cite{Bachtold2013, Wang2017Oct, Bachtold2018}, cooling \cite{OConnell2010Apr}, transfer of quantum states to mechanical motion \cite{Lehnert2013,Lehnert2017} and in the exploration of the foundation of quantum mechanics \cite{Marquardt2014Review}.
The main reason for these promising features lies in the strong back-action of a single photon or electron on the mechanical motion. Unprecedented control over quantum states is then available and macroscopic quantum states can be created allowing for foundational tests of quantum mechanics.
Recent proposals suggest that work extraction at the nanoscale is possible in the ultrastrong coupling regime \cite{Elouard_2015, Monsel_2018}, as well as the study of fluctuation theorems \cite{Monsel_2018} and study of systems far from equilibrium \cite{wachtler2019proposal, wachtler2019stochastic}. 

Among the large variety of optomechanical and electromechanical platforms developed \cite{bachtold2022Review,Treutlein2009,Clerk2010,Treutlein2010,Lukin2009,Lukin2012,Steele2012Super,treutlein2014,Sillanpaa2015,steele2009,lassagne2009coupling}, the ultrastrong coupling between quantum states and mechanical motion is within reach only for a few, including superconducting circuits ($ \gm /\omegam\simeq0.35$)\cite{Pirkkalainen2013Feb}, NV centers embedded in semiconducting nanowires under a magnetic field gradient ($ \gm /\omegam\simeq0.1$)\cite{Arcizet2011Nov}, and quantum dots in semiconducting nanowires, for which strain is the coupling mechanism ($ \gm /\omegam\simeq0.85$)\cite{Yeo2014,Poizat2020}. Theoretical proposals indicate that the ultrastrong coupling could be reached in SQUIDs with a mechanical compliant segment \cite{Blencowe2016,Blanter2017,Laird2018Displacemon,Steele2020}, single atoms in a cavity \cite{Chang2018exploring,Chang2018reaching} or Cooper pair boxes \cite{Sillanpaa2014,Nation2014,Hakonen2020}.

\begin{figure}[b]
 \centering
 \includegraphics{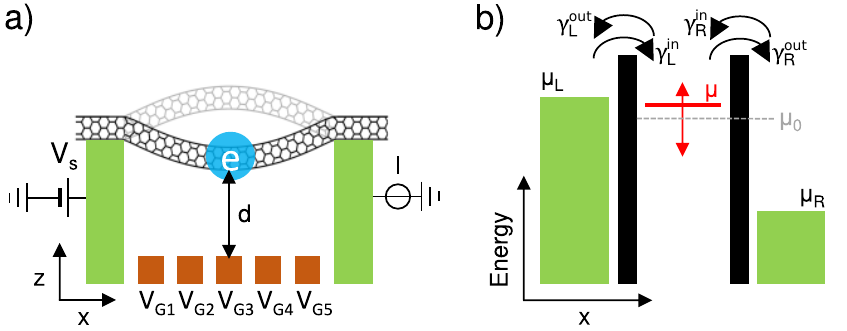}
 \caption{
     (a) Schematic of the device. A carbon nanotube is suspended between two metallic reservoirs and over an array of gate electrodes to which we apply gate voltages $V_\mathrm{G1-G5}$. A bias voltage $\Vs$ drives a current $I$ through the nanotube, within which a quantum dot is electrostatically defined. 
     The single-electron tunneling through the quantum dot couples to the nanotube's motion. The coupling strength depends on the distance $d$ between the quantum dot and the gate electrodes.
     (b) Schematic diagram of the electrochemical potential levels of a quantum dot. The left (L) and right (R) tunnel rates from the reservoirs to the quantum dot are indicated $\gL^\mathrm{\In}$, $\gL^{\Out}$, $\gR^\mathrm{\In}$ and $\gR^{\Out}$. The electrochemical potentials of left and right contacts are $\muL$ and $\muR$ respectively, and their energy difference defines a bias window. When the carbon nanotube vibrates, the electrochemical potential of the quantum dot $\upmu$ shifts with respect to a reference electrochemical potential $\upmu_0$, which is controlled by the gate electrodes.  The red arrow symbolizes the change in $\upmu$ caused by the nanotube's motion.}
 \label{fig:Device}
\end{figure}

Quantum dots electrostatically defined in fully-suspended carbon nanotube devices (Fig.~\ref{fig:Device}(a)) offer a high-degree of control over the confinement potential \cite{Ilani2014}. The mechanical properties of carbon nanotubes are also exceptional; comparatively large zero-point motion, quality factors as high as 5 million \cite{Bachtold2014} and mechanical frequencies up to 39 GHz \cite{Laird2012}. 

When the carbon nanotube is in motion, its displacement changes the distance between the carbon nanotube and the gate electrodes. The quantum dot is thus capacitively coupled to the nanotube's motion. The first evidence of this effect was the observation that single-electron tunneling creates periodic modulations of the mechanical resonance frequency \cite{Paul2002,steele2009,Bachtold2009,Huttel2010,Steele2012,Ilani2014}. These modulations of the mechanical resonance frequency, also called softening, are a signature of the electromechanical coupling between charge states and mechanical motion \cite{Pistolesi2015, Pistolesi2016}. This coupling allowed for the realization of coherent mechanical oscillators driven by single-electron tunneling~\cite{Laird2019}, cooling of the mechanical motion~\cite{Bachtold2019}, and probing of electronic tunnel rates~\cite{Ilani2019}. Fully-suspended carbon nanotube devices in the ultrastrong coupling regime have been proposed for the realization of nanomechanical qubits~\cite{Bachtold2020}. A recent study has demonstrated so called deep-strong coupling~\cite{Solano2019}, which is the equivalent of ultrastrong coupling between a carbon nanotube quantum dot and a THz resonator~\cite{Takis2021}. Until now, a careful experimental estimation of the electromechanical coupling strength that carbon nanotube devices can offer was still missing.

In this work we show that the electromechanical coupling in fully-suspended carbon nanotube devices can reach the ultrastrong coupling regime and that it presents one of the highest coupling ratios reported so far; $\gm /\omegam\simeq 2.72$. We obtain this ratio using two independent approaches. We measure the periodic modulations of the mechanical resonance frequency resulting from single-electron tunneling in our experiment and model it using a rate equation model. We also simulate the quantum dot energy levels as the carbon nanotube position changes in the plane of motion. Both approaches lead to similar conclusions and converge to the same quantitative value of $\gm$. The observed coupling ratios can be improved further by adapting the geometry of the device.

\section{System and electromechanical model}\label{sec:model}

We focus on a carbon nanotube device with a suspended segment of approximately 800~nm [see Fig.~\ref{fig:Device}(a)]. The quantum dot is defined in the nanotube through a combination of Schottky barriers at the contacts and the voltages applied to five gate electrodes (labeled V$_{G1–G5}$) beneath the nanotube. These gate electrodes are also used to actuate the nanotube's motion~\cite{Laird2016,Laird2018,Laird2019}. A current $I$ is driven by a bias voltage V$_s$. All experiments are performed at 40~mK.

To model the interplay between the single-electron transport through the quantum dot and the nanotube's mechanical motion in this device, we use rate equations. First, we describe the electron transport through the device. Applying a bias voltage $\Vs$ between the source (left) and drain (right) reservoirs opens up an energy window $e\Vs = \muL-\muR$, where $e$ is the charge of an electron, and $\muL$ and $\muR$ are the electrochemical potentials of the left and right reservoirs, respectively. If within this energy window, which we will refer to as bias window, there is an electrochemical potential level $\upmu$ corresponding to a transition that involves the charge state of the quantum dot, electrons can tunnel from one reservoir onto the quantum dot and off to the other reservoir.

We calculate the current $I$ as a function of the quantum dot electrochemical potential $\upmu$. The quantum dot is weakly coupled to left and right reservoirs [see Fig.~\ref{fig:Device}(b)], and this coupling is parameterized by four effective tunnel rates; tunneling from the left/right reservoir to the quantum dot ($\g_\LR^\In(\upmu)$) and tunneling from the quantum dot to the left/right reservoir ($\g_\LR^\Out(\upmu)$)~\cite{Steele2012}. These effective tunnel rates correspond to the product of the left/right tunnel barrier rates ($\G_\LR$) and the overlap between the density of states of the quantum dot and left/right reservoirs, $\rho_\LR(\upmu)$, i.e.
\begin{subequations}\label{eq:tunnel rates}
\begin{align}
  &\g_\LR^\In(\upmu)=\G_\LR ~\rho_\LR(\upmu),\\
  &\g_\LR^\Out(\upmu)=\G_\LR \left  (1-\rho_\LR(\upmu)\right).
\end{align}
\end{subequations}
The tunneling through the quantum dot occurs at a rate $\Gb= \sum_{\upmu=L,R} (\g_{\upmu}^{\In} + \g_{\upmu}^{\Out}) =\GL +\GR$. As we will show later, $\Gb/2\pi$ is of the order of $100~\text{GHz}$, and thus $\hbar\Gb\gg k_\text{B}T$ for sub-Kelvin temperatures, with $k_\text{B}$ the Boltzmann constant. In this regime, we find~\cite{Steele2012,beenakker1991}
\begin{equation}
\rho_\LR(\upmu)=\frac{1}{2}+\frac{1}{\pi}\arctan \left( \frac{2(\upmu_\LR-\upmu)}{\hbar\Gb} \right ).\label{eq:rho}
\end{equation}
We can thus express the current flowing through the quantum dot as
\begin{equation}
I(\upmu) = e  \frac{\gL^\In(\upmu)  \gR^\Out(\upmu)-\gR^\In(\upmu)  \gL^\Out(\upmu)}{\Gb}.
\label{equ:current}
\end{equation}

We now examine how the mechanical motion affects the electron transport. As the carbon nanotube moves, its displacement $z$ in the vertical direction changes the capacitance between the gate electrodes and the quantum dot. This leads to a change in $\upmu$ proportional to the electromechanical coupling constant $\gm$ at the first order in the displacement [see Appendix \ref{appendix:optomechanics}],
\begin{equation}
  \upmu(z)\simeq\upmu_0+ \hbar \gm \frac{z}{\zpm},
\label{eq:epsilon}
\end{equation}
where $\zpm=\sqrt{\hbar/2 m \omegam}$ is the zero point motion, with $m$ the nanotube's mass (see Appendix \ref{appendix:mass} for details on the estimation of the carbon nanotube's mass), and $\upmu_0$ is the electrochemical potential of the quantum dot for a carbon nanotube displacement equal to 0, i.e. at $z=0$. We can control $\upmu_0$ with the applied gate voltages.  

The change in $\upmu$ caused by the nanotube's motion produces a change in the average population of the quantum dot. This change can be considered adiabatic if $\Gb \gg \omegam$. This means that, on the timescales corresponding to the mechanical motion, the average population instantaneously reaches a steady-state and is purely defined by the position of the carbon nanotube. In this regime, electron–vibron coupling mechanisms such as the Franck–Condon blockade \cite{Andreev2006,leturcq2009,mariani2009} are negligible. In this case, we find that the relative average occupation of the quantum dot with reference to a fixed charge state is
\begin{equation}
  p(\upmu(z)) = \frac{\gL^\In(\upmu(z))+\gR ^\In(\upmu(z))}{\Gb}.
  \label{equ:pop}
\end{equation}
Note that $p$ is a number between 0 and 1.

The mechanical motion is in turn affected by the electron transport. Variations of $p$ cause the reduction of the mechanical resonance frequency that is considered a signature of strong electromechanical coupling in nanotube mechanical resonators~\cite{Paul2002,steele2009,Bachtold2009,Huttel2010,Steele2012,Ilani2014}. The effective resonance frequency $\weff(\upmu_0)/2\pi$, lower than $\omegam/2\pi$, is observed when $\upmu$ varies within the bias window [$\muR, \muL$]. This interplay between single-electron transport and mechanical motion can be explored further by writing the equation of motion that models the carbon nanotube displacement [see Appendix \ref{appendix:optomechanics}],

\begin{equation}
  \ddot{z} + \omegam^2\left[z + 2 p(\upmu(z)) \frac{\gm}{\omegam}\zpm\right] = 0. \label{eq:eq_of_motion}
\end{equation}
Since carbon nanotube devices exhibit high quality factors, we neglect the mechanical damping over a few mechanical periods.

\begin{figure}[t]
    \centering
    \includegraphics[width=\columnwidth]{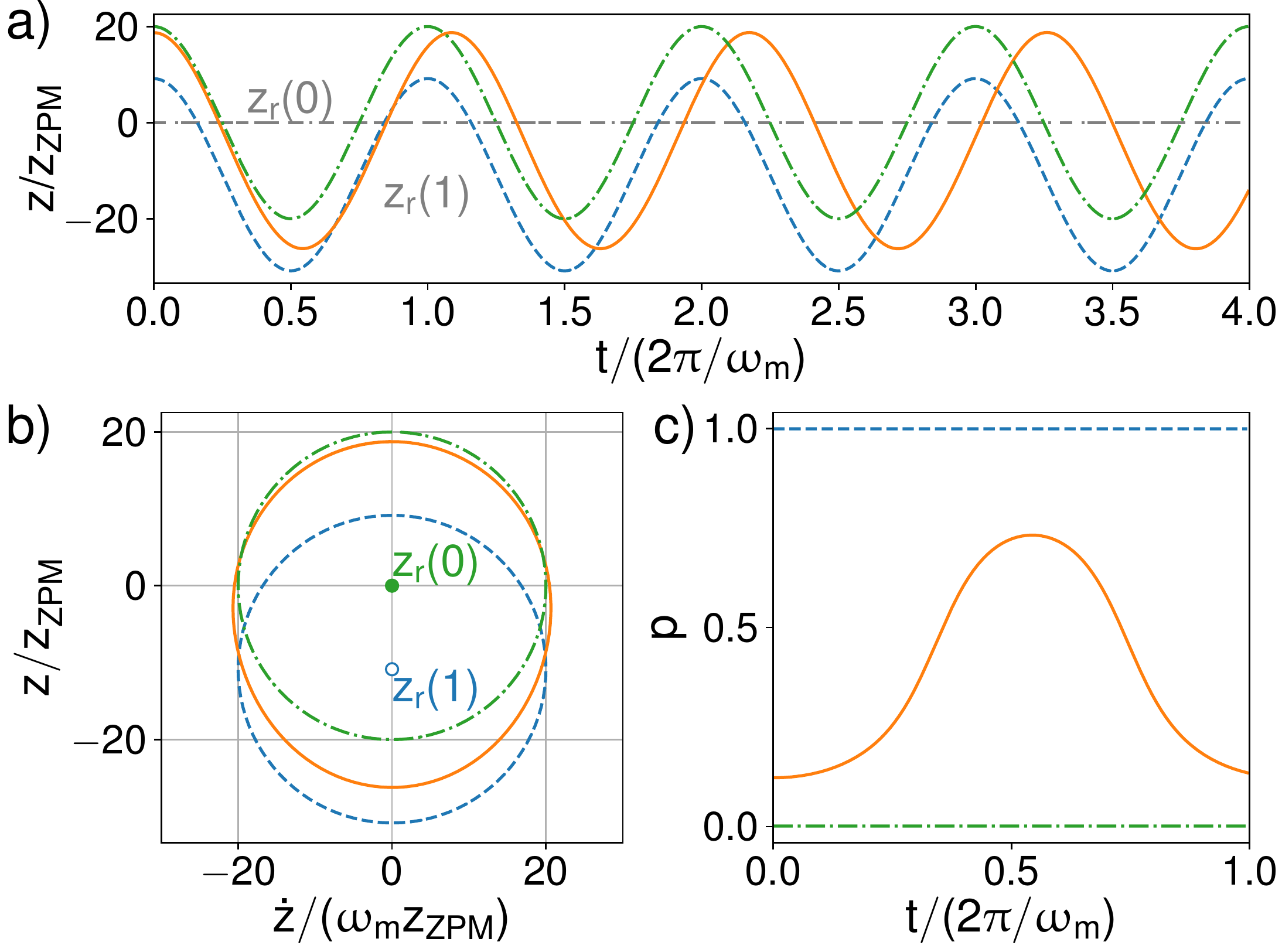}
    \caption{
        Time evolution obtained by numerically integrating the equation of motion~\eqref{eq:eq_of_motion}, starting from the initial conditions $z(0) = z_0 + z_r(p(\upmu(z_0)))$ and $\dot{z}(0)=0$, with $z_0/\zpm = 20$, for three different values of $\upmu_0$: well above the bias window ($\upmu_0 \gg \muL$, dash-dotted green lines), within the bias window ($\upmu_0 = 0$, solid orange lines) and well below the bias window ($\upmu_0 \ll \muR$, dashed blue lines). This choice of $z(0)$ results in identical amplitudes for the cases $\upmu_0 \gg \muL$ and $\upmu_0 \ll \muR$. 
        (a) Nanotube's displacement $z$ as a function of time over four mechanical periods.  Grey lines indicate the nanotube's rest position for $p = 0$ and $p = 1$: $z_r(0) = 0$ and $z_r(1) = - 2( \gm/\omegam)\zpm$.
        (b) Corresponding phase space trajectories over one mechanical period. The dots indicate the resonator's rest position for $p = 0$ and $p = 1$: $z_r(0)$ (green point) and $z_r(1)$ (blue circle).
        (c) Population of the quantum dot as a function of time over one mechanical period.
       These simulations, use the parameters extracted from the experiment (see main text); $\omegam/2\pi=294.5$ MHz, $\GL/2\pi = 1.0$ GHz, $\GR/2\pi = 40$ GHz, 
       and  $\zpm = 0.68$ pm [see Appendix \ref{appendix:characterization}]. In the plots, the value of $\gm/2\pi$ was exaggerated by a factor 2 for visual clarity; $\gm/2\pi = 1.6$ GHz while the true value is $\gm/2\pi = 0.8$ GHz.
    }
    \label{fig:Model}
\end{figure}

The combination of Eqs.~\eqref{equ:pop} and ~\eqref{eq:eq_of_motion} makes explicit that $p$ can change within a mechanical oscillation, since $p$ depends on $\upmu$, and $\upmu$ depends on $z$ (Eq.~\eqref{eq:epsilon}), which is a function of time. Considering that $\upmu$ has a weak dependence on $z$, the rest position of the resonator $z_r(p) = - 2 p(\upmu)(\gm/\omegam)\zpm$, is obtained for $\ddot{z}=0$.
Figures~\ref{fig:Model}(a-c) show the nanotube's displacement $z$ and dot population $p$ as a function of time, as well as the corresponding trajectories in phase space, obtained by solving Eq.~\eqref{eq:eq_of_motion} numerically for different values of $\upmu_0$. When $\upmu_0$ is far above the bias window ($\upmu_0 \gg \muL$), $p(\upmu) = 0$ and the resonator rest position is $z_r(0) = 0$ (dash-dotted green line). Conversely, when $\upmu_0$ is far below the bias window ($\upmu_0 \ll \muR$), the population is $p(\upmu) = 1$ and the nanotube's rest position is $z_r(1) = - 2(\gm/\omegam)\zpm$ (dashed blue line). But when $\upmu_0$ is within the bias window, $p(\upmu)$ varies between 0 and 1, i.e. $0 \leq p(\upmu) \leq 1$, and the nanotube's rest position satisfies $0\geq z_r(p) \geq - 2( \gm/\omegam)\zpm$ (solid orange line). The nanotube's motion follows a trajectory in phase space at constant angular velocity $\omegam$ but the rest position $(0, z_r(p))$ shifts with $p(t)$, making the trajectory elliptical instead of circular [see Fig.~\ref{fig:Model}(b)). As a result, when $p$ is not constant (solid orange line), the period can exceed $2\pi/\omegam$, leading to a reduction of the effective mechanical resonance frequency  $\weff(\upmu_0)/2\pi$, evident in Fig.~\ref{fig:Model}(a).

Under the approximation of small displacements, the effective  resonance frequency can be estimated from Eq.~\eqref{eq:eq_of_motion}. Linearizing $p(\upmu)$, we obtain
\begin{equation}\label{eq:p_lin}
  p(\upmu) 
  = p(\upmu_0)+(\upmu - \upmu_0)\left.\frac{\partial p}{\partial \upmu}\right\vert_{\upmu_0}.
\end{equation}
We introduce this expression in Eq.~\eqref{eq:eq_of_motion} and use Eq.~\eqref{eq:epsilon} to rewrite the equation of motion as follows
\begin{equation}
  \!\ddot{z} + \left[\omegam^2 
  + 2\gm^2 \hbar \omegam \left.\pdv{p}{\upmu}\right\vert_{\upmu_0}\!\right] z
  = - 2 p(\upmu_0) \gm\omegam\zpm. \!
  \label{eq:eq_of_motion_1st_order}
\end{equation}
Thus the effective resonance frequency is
\begin{equation}
  \weff(\upmu_0) = \sqrt{\omegam^2 
      + 2\gm^2 \hbar \omegam \left.\pdv{p}{\upmu}\right\vert_{\upmu_0}}.
  \label{eq:softening}
\end{equation} 
Note that $\partial p / \partial \upmu$ is negative.

\section{Experimental results}\label{sec:exp}

\begin{figure}[htb]
    \centering
    \includegraphics{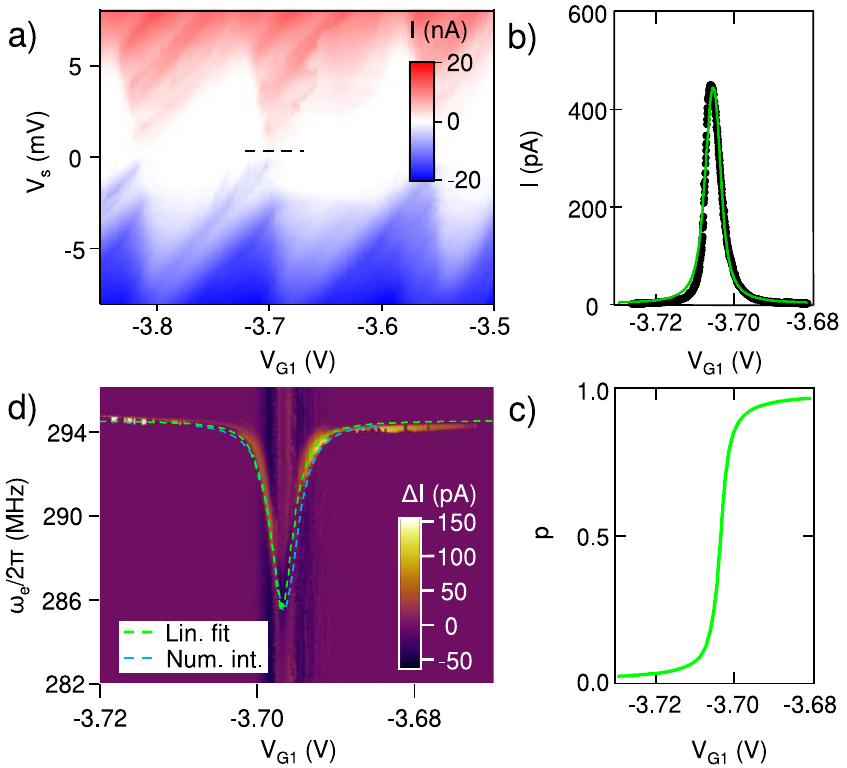}
    \caption{
        (a)  Current as a function of $\Vs$ and $V_\mathrm{G1}$ with $V_\mathrm{G2}=0$~V, $V_\mathrm{G3}=-4$~V, $V_\mathrm{G4}=0$~V, $V_\mathrm{G5}=-3.95$~V and $R_\mathrm{s}=100$~k$\Omega$. We observe Coulomb diamonds, evidencing single-electron tunneling.
        (b) Current as a function of $V_\mathrm{G1}$ for $\Vs=0.2~\text{mV}$ at the Coulomb peak indicated by the black dashed line in (a). The green line is a fit to Eq.~\eqref{equ:current} with $\GL/2\pi=1.0~\text{GHz}$ and $\GR/2\pi= 40~\text{GHz}$.
        (c) Population $p$ as a function of $V_{G5}$ computed using Eq.~\eqref{equ:pop} for $z=0$.
        (d) Current variation $\Delta I$ as function of the drive power frequency $\omega_\mathrm{e}/2\pi$ and $V_\mathrm{G1}$ at $\Vs = 0.2~\text{mV}$. In each column, the average value of the current was subtracted to highlight the mechanical resonance.
        Near the Coulomb peak, the nanotube's resonance frequency shows a dip. A small shift in $V_\mathrm{G1}$ explains the gate voltage difference between the center of the  Coulomb peak in panels (b) and (d). The green line (Lin. fit) is a fit with Eq.~\eqref{eq:softening} and the blue line (Num. int.) is the effective resonance frequency reproduced by numerical integration of the equation of motion \eqref{eq:eq_of_motion} with $\gm/2\pi=0.80~\text{GHz}$.}
    \label{fig:Experiments}
\end{figure}

To verify the validity of this prediction in our device and estimate the coupling strength $\gm$, we use gate voltages to define a single quantum dot, revealed by the Coulomb diamonds in Fig.~\ref{fig:Experiments}(a). From this measurement, we estimate the lever arm $\alpha=0.054_{-0.005}^{+0.007}$~eV/V and its uncertainty [see Appendix \ref{appendix:alpha}], which relates the variation of $\upmu_0$ with the applied gate voltages $\Delta \upmu_0=-\alpha\Delta V_\mathrm{G1}$. Measurements in Fig.~\ref{fig:Experiments}(a,b) were performed with the carbon nanotube at rest (no driven motion), and thus $\upmu$ is equal to $\upmu_0$. From a fit of a Coulomb peak using Eq.~\eqref{equ:current} (Fig.~\ref{fig:Experiments}(b)), we obtain $\GL/2\pi= 1.0 \pm 0.1~\text{GHz}$ and $\GR/2\pi= 40 \pm 5 ~\text{GHz}$. The uncertainty interval in these tunneling rates is determined by fitting the Coulomb peak with two extreme $\alpha$ values given by the uncertainty in $\alpha$. We then use Eq.~\eqref{equ:pop} to estimate $p$ for any value of $\upmu$. The resulting $p(\upmu)$ is shown in Fig.~\ref{fig:Experiments}(c).

We drive the nanotube's motion by a microwave tone at frequency $\omega_\mathrm{e}/2\pi$ and drive power $P_0=-79~\text{dBm}$ applied to gate G3 [see Fig.~\ref{fig:Device}]. The mechanical resonance causes sharp steps in $I(\omega_\mathrm{e}/2\pi)$. Numerically differentiating $I(\omega_\mathrm{e}/2\pi)$, the resonance is evident as peaks/dips in $\mathrm{d}I/\mathrm{d}\omega_\mathrm{e}$ (Fig.~\ref{fig:Experiments}(d)).
The mechanical resonance frequency drops below $\omegam/2\pi=294.5~\text{MHz}$ at values of $V_\mathrm{G1}$ for which we observed a Coulomb peak (Fig.~\ref{fig:Experiments}(b)). We fit this effective resonance frequency, $\weff$, using Eq.~\eqref{eq:softening}. Because $p(\upmu)$ is estimated from the tunnel rates and $\upmu_0$ is calculated from the lever arm $\alpha$ (Fig.~\ref{fig:Experiments}(a,b)), the coupling strength $\gm$ is the only fitting parameter. 
The resistance of the measurement circuit was taken into account by correcting the bias voltage accordingly [see Appendix \ref{appendix:bias correction}].
We find $\gm/2\pi = 0.80 \pm 0.04~\text{GHz}$ given the uncertainty over $\GL$ and $\GR$. This result leads to a coupling ratio $\gm /\omegam\simeq 2.72 \pm 0.14$. This ratio, is, to the best of our knowledge, the highest value reported among all other electromechanical platforms. We have estimated $\gm/2\pi$ for other Coulomb peaks in Appendix \ref{appendix:extra peaks}.

We have further corroborated $\weff(\upmu_0)$ by numerically integrating Eq.~\eqref{eq:eq_of_motion}. This approach does not require $p(\upmu)$ to be linearized. We estimate $\zpm = 0.68$~pm [see Appendix \ref{appendix:mass}], and considering the values of $\omegam$, $\GL$ and $\GR$ extracted from the experiment, we compute $z(t)$ for various sets of values of $\upmu_0$, $\gm$ and $z(0)$, choosing $\dot{z}(0) = 0$. We then derive $\weff$ [see Appendix \ref{appendix:numerical simualtions}] and find that $\gm/2\pi \simeq 0.80$~GHz accurately reproduces the dependence of $\weff$ with $V_\mathrm{G1}$ observed in the experiment (dashed blue line in Fig.~\ref{fig:Experiments}(d)). This result is in good agreement with the value of $\gm$ obtained from the fit to Eq.~\eqref{eq:softening}. The amplitude of motion, $z(0) \simeq 20\zpm \sim 15$~pm, is consistent with the values estimated in previous experiments \cite{Laird2018,Laird2019}. The value of $z(0)$ only significantly affects the width of the dip in the resonance frequency when $z(0)/\zpm$ is larger than $\Gb/\gm$, i.e $z(0)/\zpm \gtrsim  50$ [see Appendix \ref{appendix:mass}]. We thus confirm that the small displacement limit (Eq.~\eqref{eq:p_lin}) applies to our experiments.

\section{Semiclassical electrostatic model}\label{sec:electrostatics}

We now compare these results with a semiclassical numerical approximation.
We calculate the single-particle energy levels of the dot, $\varepsilon_n(z)$, $n = 0,1,...$. In this case, $\varepsilon_n(z)$ is the contribution to the charging energy, $\upmu$, which depends on position $z$ for the gate voltage configuration of the experiment. The occupied energy levels will only impose a constant force on the oscillator.
In this case, the value of $\gm$ can be estimated from Eq.~$\eqref{eq:epsilon}$ as
\begin{equation}
    \gm= \dfrac{\zpm}{\hbar}\dfrac{d\varepsilon_{n} }{dz}.
\end{equation}

We compute the levels $\varepsilon_n(z)$ solving explicitly the electric potential field in the plane of motion, $V(z,x)$, using a finite difference method [see Appendix \ref{appendix:electrostatics}]. Then, the dot energy levels can be obtained from $V(z,x)$ using the Bohr-Sommerfeld equation \cite{nazarov2009quantum},
\begin{equation}
  \oint \sqrt{2m_e\left(\varepsilon_n(d) - eV(d,x)\right)}dx = 2\pi \hbar\left(n+\frac{1}{2}\right).
  \label{eq:bohr}
\end{equation}
The integral is calculated along an horizontal line at height $d$ from the gates representing the classical path of the electrons. $m_e$ is the electron mass.
\begin{figure}[t]
  \centering
  \includegraphics[scale = .35]{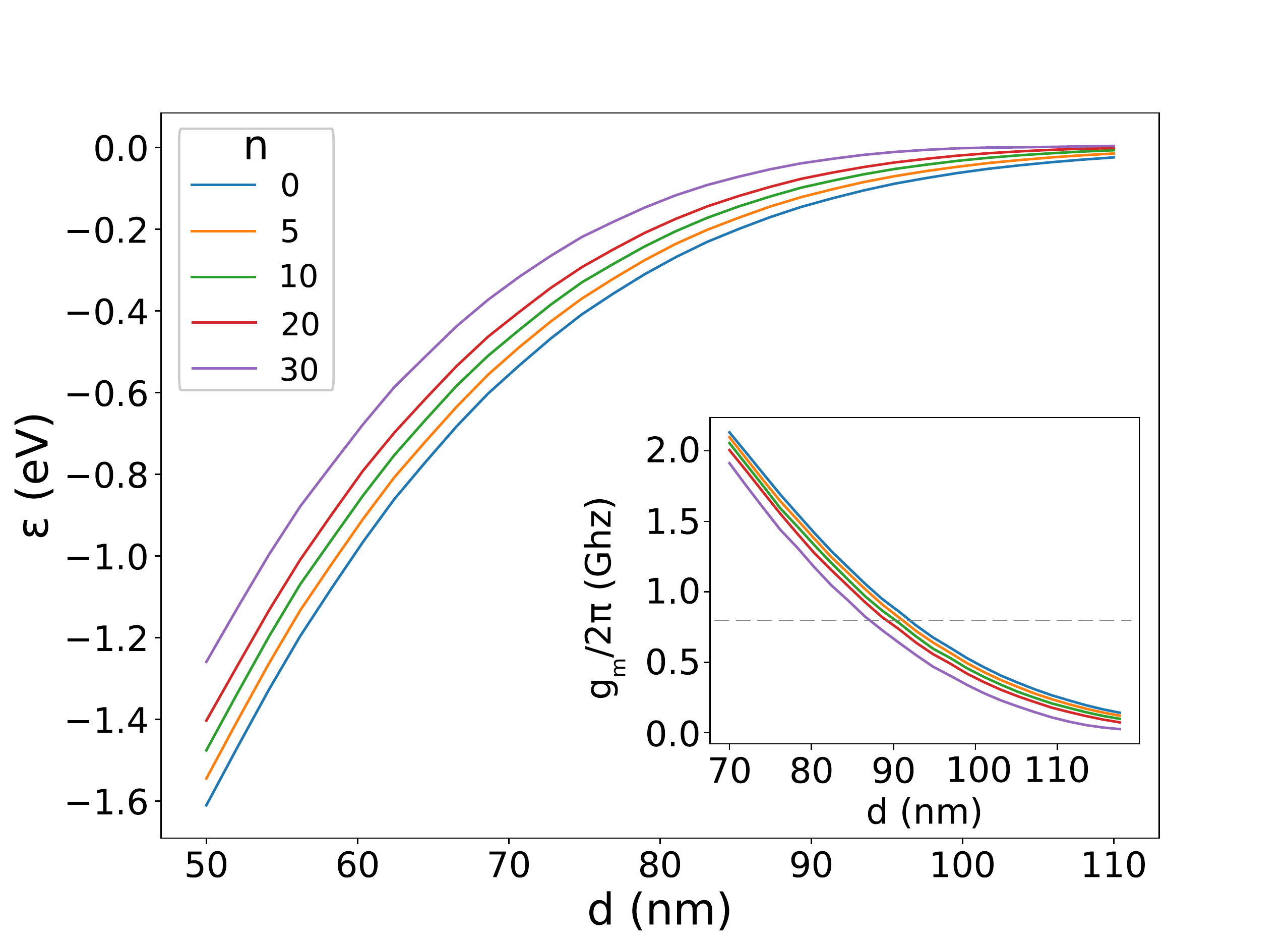}
  \caption{Energy levels of the quantum dot with $n = 0, 5, 10, 20, 30$ as a function of $d$ obtained from Eq.~\eqref{eq:bohr}. (Inset) Values of $\gm$ obtained from the dependence of the energy levels with $d$. The dash line represent the value $\gm/2\pi=0.80$~GHz obtained from the experiment.}
  \label{fig:electromagnetic}
\end{figure} 
Fig.~\ref{fig:electromagnetic} shows $\varepsilon$ for different values of $n$ as a function of the distance $d$. The values of $\gm$ extracted for different values of $d$ are displayed in Fig.~\ref{fig:electromagnetic}(inset). We find a value  $\gm/2\pi \approx 0.80$~GHz for $d = 90$~nm, a distance which is consistent with the geometry of our device [see Appendix \ref{appendix:electrostatics}] considering the deformation of the nanotube. The value of $\gm$ decreases slightly with the quantum level index $n$ and as a function of $d$, setting the range of possibilities for our platform.

\section{Conclusion}

To conclude, we have found that fully-suspended carbon nanotube devices can reach ultrastrong coupling $\gm/2\pi \approx 0.80 \pm 0.04$~\text{GHz} between single-electron transport and mechanical motion, 
leading to a coupling ratio of $\gm /\omegam\simeq 2.72 \pm 0.14$, a value that exceeds that obtained with any other electromechanical platform.
We have quantified the coupling strength by using rate equations to model the reduction of mechanical resonance frequency observed in our experiments. We separately  confirmed the resulting coupling strength  with electrostatic simulations based on Bohr-Sommerfeld equations. From these simulations, we extrapolate that this coupling could be enhanced by reducing the distance between the carbon nanotubes and the gates and/or the number of charges in the quantum dot. Using our model to fit measurements from similar suspended carbon nanotube devices (\cite{Steele2012} and \cite{Huttel2010}), we concluded that the ultrastrong coupling regime was present, but went unnoticed. We obtained ratios $\gm/\omegam$ of 1.7 and 1.25, respectively [see Appendix \ref{appendix:other devices}]. This finding suggests that the ultrastrong coupling regime is standard in this type of devices. It allows for an ambitious suite of experiments, ranging from nanomechanical qubits to information to work conversion at the nanoscale.

\begin{acknowledgments}
We acknowledge useful discussions with M. Woolley and F. Pistolesi and thanks Serkan Kaya for his help in the fabrication of the device. This research was supported by grant number FQXi-IAF19-01 from the Foundational Questions Institute Fund, a donor advised fund of Silicon Valley Community Foundation. 
NA acknowledges the support from the Royal Society, EPSRC Platform Grant (grant number EP/R029229/1), from the European Research Council (ERC) under the European Union's Horizon 2020 research and innovation programme (grant agreement number 948932), and from Templeton World Charity Foundation.
AA acknowledges the support of the Foundational Questions Institute Fund (grant number FQXi-IAF19-05), the Templeton World Charity Foundation, Inc (grant number TWCF0338) and the ANR Research Collaborative Project ``Qu-DICE" (grant number ANR-PRC-CES47).
JT and JMRP acknowledge financial support from the Spanish Government (Grant Contract, FIS-2017-83706-R).
JA acknowledges support from EPSRC (grant number EP/R045577/1) and the Royal Society. 
JM acknowledges funding from the Vetenskapsr\r{a}det, Swedish VR (project number 2018-05061).
\end{acknowledgments}

\appendix

\section{Simplified electromechanical model in the adiabatic regime}\label{appendix:optomechanics}
In this appendix, we give more details about the electromechanical model presented in Sec.~\ref{sec:model}.

We focus here on a single level of the quantum dot, the one inside or closest to the bias window, assuming that there is at most one level inside it (as represented in Fig.~\ref{fig:Device}(b)).
The quantum dot is capacitively coupled to the gates and the effective capacitance depends on the distance between the quantum dot and the gates. Therefore, the vertical motion of the carbon nanotube (CNT) changes the electrochemical potential $\upmu$ of the quantum dot level. At the first order in $z$, we have
\begin{equation}
    \upmu(z) \simeq \upmu_0+ \left.\pdv{\upmu}{z}\right\vert_{z=0} z,
\end{equation}
where $z=0$ corresponds to the rest position of the carbon nanotube when the quantum dot level is empty. Like for optomechanical systems \cite{Marquardt2014Review}, we define from the above expression the electromechanical coupling strength 
\begin{equation}
    \gm = \frac{1}{\hbar}\left.\pdv{\upmu}{z}\right\vert_{z=0} \zpm,
\end{equation}
and obtain the expression of $\upmu(z)$ given by Eq.~\eqref{eq:epsilon}. \\

So the Hamiltonian describing this simplified model of the electromechanical system is 
\begin{eqnarray}
    H = \left(\upmu_0 + \hbar \gm \frac{\hat{z}}{\zpm}\right)  \hat{n} + \hbar \omegam \hat{b}^\dagger \hat{b},
\end{eqnarray}
where $\hat{b}$ is the annihilation operator of the considered mechanical mode and $\hat{n}$ the occupation of the quantum dot level. The interaction part of the Hamiltonian therefore writes
\begin{equation}
    H_\text{int} =  \hbar \gm \frac{\hat{z}}{\zpm}\hat{n},
\end{equation} 
which corresponds to an electromechanical force 
\begin{equation}\label{F_em}
    \hat{F} = -\dv{H_\text{int}}{\hat{z}} = -\frac{\hbar \gm \hat{n}}{\zpm} 
\end{equation}
applied on the resonator.\\

In addition, electrons tunnel in and out the quantum dot with rates $\g_\LR^\In(\upmu)$ and $\g_\LR^\Out(\upmu)$ [see Eqs.~\eqref{eq:tunnel rates}] and the mechanical resonators undergoes damping at rate $\g_\mathrm{m}$. Our device operates in the semi-classical regime (large phonon number in the resonator) where there is no entanglement between the quantum dot and resonator and no coherences inside the quantum dot. Furthermore, the relevant time scales for the tunneling events $1/\G_\LR$ are orders of magnitude shorter than the mechanical dynamics [see Table~\ref{table:params}]. Therefore, we make the adiabatic approximation, namely we consider that the population $p = \ev{\hat{n}}$ of the quantum dot is always the equilibrium one [Eq.~\eqref{equ:pop}],
and instantaneously follows the variations of $z = \ev{\hat{z}}$. The time evolution of the position of the resonator is described by the classical equation of motion
\begin{equation}
    \ddot{z} + \gm \dot{z} + \omegam^2 z =  \frac{\mean{\hat{F}}}{m},
\end{equation}
where $m$ is the mass of the resonator.  This equation is consistent with the results from Ref.~\cite{Pistolesi2015}. In the following, we will consider only a few mechanical periods and thus neglect the mechanical damping due to the high quality factor $Q_\mathrm{m} = \omegam/\g_\mathrm{m}$ [see Table \ref{table:params}]. Using Eq.~\eqref{F_em} and the expression of the zero-point motion fluctuation $\zpm = \sqrt{\hbar/2m\omegam}$, we obtain the equation of motion \eqref{eq:eq_of_motion}.

\section{Characterization of the experimental device}\label{appendix:characterization}
In this appendix, we describe the suspended carbon nanotube device we used in the experiment and explain how we determined its characteristics.

\subsection{Carbon nanotube device}\label{appendix:device}
\begin{figure}[htb!]
    \centering
    \includegraphics[width=\linewidth]{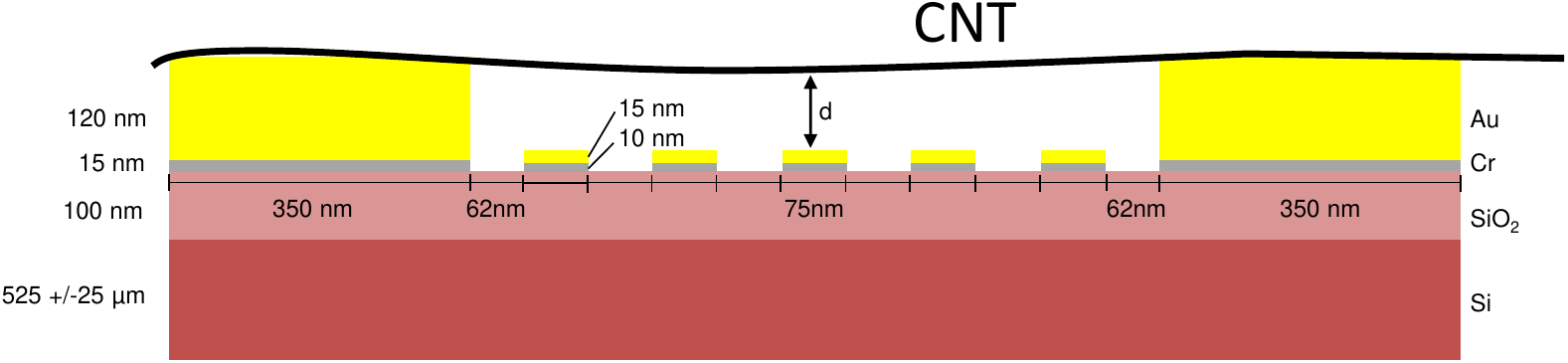}
    \caption{Schematic of the device.}
    \label{fig:SuppDevice}
\end{figure}

The suspended carbon nanotube device is similar to the one presented in \cite{Laird2016,Laird2018,Laird2019}. We fabricated chips from high resistance Si/SiO$_\text{2}$ substrate by patterning Au/Cr electrodes with Ebeam lithography. The carbon nanotubes are grown by CVD on a separate quartz substrate using a nanoparticles of Al$_\text{2}$O$_\text{3}$, Fe(NO$_\text{3}$) and MoO$_\text{2}$(acac)2 as catalyst and mechanically transferred to the chip. Figure \ref{fig:SuppDevice} display a schematic of the device respecting geometric the proportions.

\begin{table}[htb]
    
    \begin{tabularx}{\linewidth}{Xll}
        \hline\hline
         Parameter & Name & Value\\
         \hline
        Bias voltage &  $V_s$ & 0.2~mV\\
        Left tunneling rate  & $\GL/2\pi$ &$1.0 \pm 0.1~\text{GHz}$\\
        Right tunneling rate  & $\GR/2\pi$ &$ 40 \pm 5~\text{GHz}$\\
        Lifetime broadening &  $\Gb/2\pi$ &$41 \pm 5~\text{GHz}$\\
        Bare mechanical frequency &  $\omegam/2\pi$ & 294.5~MHz\\
        Zero point motion fluctuation &  $\zpm$ & $0.68 \pm 0.04~\text{pm}$ \\
        Mechanical quality factor &  $Q_\mathrm{m}$ & $\leq 2000$ \\
        Coupling strength &  $\gm/2\pi$ & $0.80\pm0.04$~GHz \\  
        \hline\hline
    \end{tabularx}
    \caption{\label{table:params}
        Parameters of the case considered in the main text.
    }
\end{table}

\subsection{Determination of the lever arm $\alpha$ and its uncertainty}\label{appendix:alpha}

\begin{figure} [h!]
    \centering
    \includegraphics{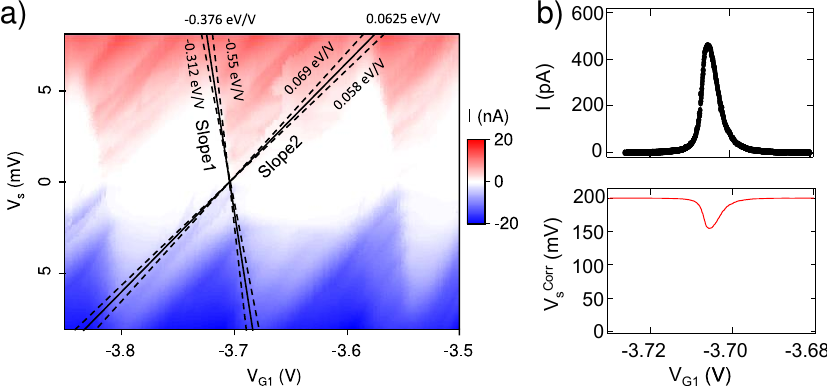}
    \caption{
        (a) Coulomb diamond from Fig.~\ref{fig:Experiments}. The two black lines follow the edge of the diamond and correspond to $\mathrm{Slope1}=-0.376_{-0.174}^{+0.064}$~eV/V and $\mathrm{Slope2}=0.0625_{-0.0045}^{+0.0065}$~eV/V. The dashed lines indicates the error in the determination of the slope.
        (b) Coulomb peak in Fig.~\ref{fig:Experiments}(b) where the measured current $I$ was smoothed to remove the noise (upper plot) and corrected voltage between source and drain contacts (lower plot).
    }
    \label{fig:alpha and biasCorr}
\end{figure}

The lever arm $\alpha=\frac{|e|C_\mathrm{G}^{l^\mathrm{dot}}}{C}$ (where $C_\mathrm{G}^{l^\mathrm{dot}}$ is the capacitance between the gate voltage $V_\mathrm{G1}$ and the dot, and $C$ the sum of the gate, source $C_\mathrm{S}$ and drain capacitances) is critical to in the estimation of the coupling strength. The lever arm can be extracted from the two slopes $\mathrm{Slope1}$ and $\mathrm{Slope2}$ of the Coulomb diamond (Fig.~\ref{fig:alpha and biasCorr}(a))~\cite{hanson2007}
\begin{equation}
    \mathrm{Slope1} = -\frac{|e|C_\mathrm{G}^{l^\mathrm{dot}}}{C-C_\mathrm{S}}, \quad \mathrm{Slope2} = \frac{|e|C_\mathrm{G}^{l^\mathrm{dot}}}{C_\mathrm{S}}.
    \label{eq:slope}
\end{equation}
Combining the two expressions of Eq.~\eqref{eq:slope}, we obtain
\begin{equation}
    \alpha = \frac{\mathrm{Slope1} \times \mathrm{Slope2}}{\mathrm{Slope1}-\mathrm{Slope2}}
\end{equation}

From Fig.~\ref{fig:alpha and biasCorr}(a), we deduce the two slopes $\mathrm{Slope1}=-0.376_{-0.174}^{+0.064}$~eV/V and $\mathrm{Slope2}=0.0625_{-0.0045}^{+0.0065}$~eV/V, resulting in a lever arm:
\begin{equation}
    \alpha = 0.054_{-0.005}^{+0.007}~\mathrm{eV/V}.
\end{equation}

\subsection{Corrections to the bias voltage}\label{appendix:bias correction}

The internal resistance $R_\mathrm{s} = 100$~k$\Omega$ of the IV converter become a significant fraction of the total resistance of the circuit when the device is tuned in a Coulomb peak. It is therefore necessary to introduce a corrected bias voltage,
    \begin{equation}
        \Vs^\text{Corr}(V_\mathrm{G1})=\Vs-I(V_\mathrm{G1})R_\mathrm{s}.
    \end{equation}

The resulted $\Vs^\text{Corr}(V_\mathrm{G1})$ is plotted in Fig.~\ref{fig:alpha and biasCorr}(b). The corrected bias voltage was used to fit the Coulomb peak in Fig.~\ref{fig:Experiments}(b) and the mechanical resonance frequency in Fig.~\ref{fig:Experiments}(d), which impact the estimation of $\gm$.

\subsection{Estimation of the carbon nanotube's mass}\label{appendix:mass}

\begin{figure}[h!]
    \centering
    \includegraphics[width=\linewidth]{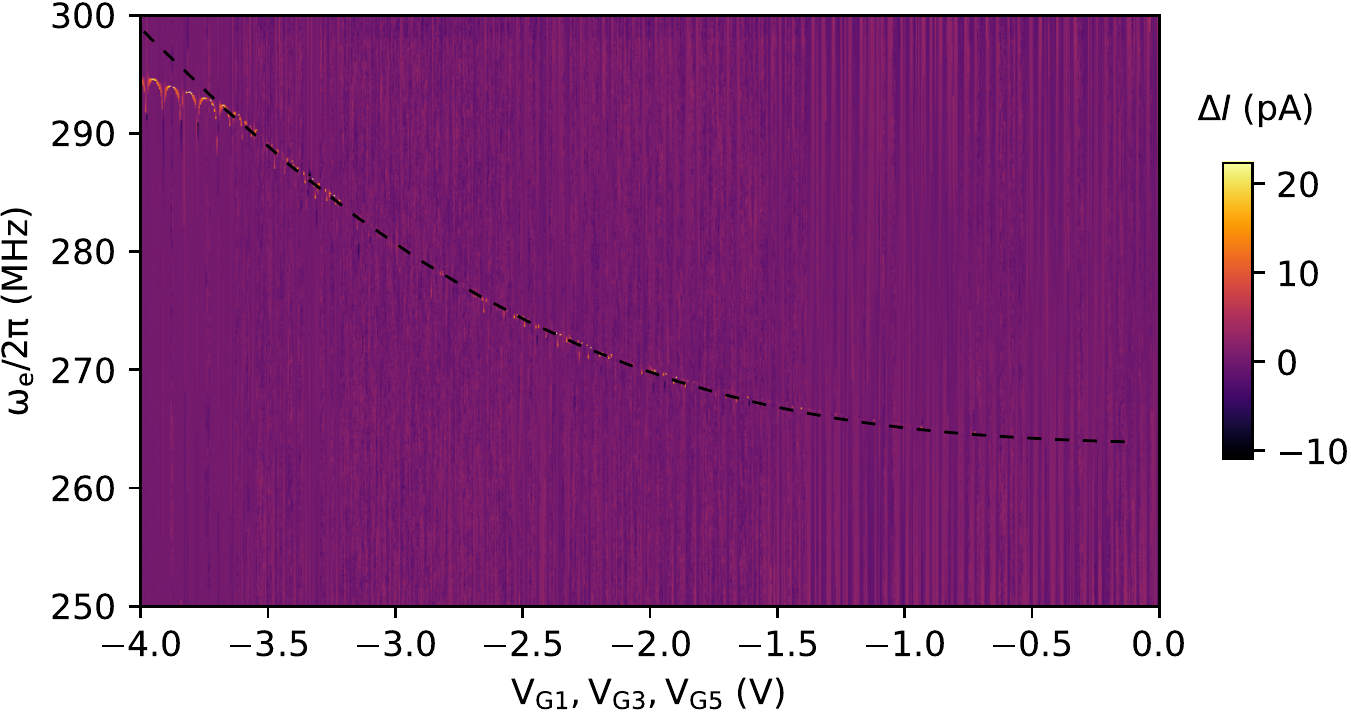}
    \caption{Mechanical resonance frequency observed for a large 
        sweep of gate voltages
        $V_\mathrm{G} = \{V_\mathrm{G1}, V_\mathrm{G3}, V_\mathrm{G5}\}$
        while driving the CNT with a microwave tone at frequency $\omega_\mathrm{e}/2\pi$
        in a similar manner as  Fig.~\ref{fig:Experiments}(d). The mechanical resonance
        frequency $\omegam/2\pi$ is fitted (black dashed line) by self-consistently
        solving for the eigenmodes of Eq.~\eqref{eq:fitm:modeeq} together with 
        Eq.~\eqref{eq:fitm:Feq} and Eq.~\eqref{eq:fitm:Teq}.
        The obtained fitting parameters are $L = 936\pm10\,\mathrm{pm}$,
        $r = 3.9\pm0.2\,\mathrm{pm}$, $V_0 = 0.82\pm0.07\,\mathrm{V}$,
        $T_r = 0.7\pm0.2\,\mathrm{nN}$.}
    \label{fig:mass}
\end{figure}

In the following we estimate the mass $m$ of the CNT and its zero point motion $\zpm$ from the dependence of the mechanical resonance frequency $\omegam/2\pi$  with gate voltage~\cite{sazanova2004,poot2007,wu2011}. We measure the change in current as a function of $\omegam$ (Fig.~\ref{fig:mass}) while sweeping three gate voltages $V_\mathrm{G}=\{V_\mathrm{G1},V_\mathrm{G3},V_\mathrm{G5}\}$ ($V_\mathrm{G2}$ and $V_\mathrm{G4}$ showed leakage currents during the experiment). We observe the increase of $\omegam$ when $V_\mathrm{G}$ become more negative until $-3.6$~V, where the CNT enters the strong bending regime~\cite{VanderZan2003,poot2007}.

\begin{table}[htb]
    \setlength{\tabcolsep}{10pt}
    \begin{tabular}{ll}
        \hline\hline
        Parameter & Estimated value \\
        \hline
        $d$     & 100~nm \\
        $L$     & $936\pm10$~nm \\
        $r$     & $3.9\pm0.2$~nm \\
        $\rho$  & 1350\,$\mathrm{kg}/\mathrm{m}^3$\\
        $E$     & 1.25\,TPa\\
        $T_r$   & $0.7\pm0.2$~nN\\
        $V_0$   & $0.82\pm0.07$~V \\
        $C_\mathrm{G}$ & $12.9\pm0.2$~aF \\
        $\partial C_\mathrm{G}/{\partial z}$ & $29.0\pm0.8$~pF/m \\
        $m$     & $61\pm6$~ag \\
        $\zpm$  & $0.68\pm0.04$~pm \\
        \hline\hline
    \end{tabular}
    \caption{\label{table:CalcMass}
        Parameters of the estimation of the carbon nanotube's mass and zero point motion $\zpm$.
    }
\end{table}

To fit the mechanical frequency, we make use of the continuum model developed
in~\cite{poot2007} and~\cite{witkamp2009thesis} to describe the bending modes of
a CNT. The displacement $z$ as a function of time and the position $x$ along 
the tube axis is modeled by the equation
\begin{equation} \label{eq:fitm:modeeq}
    \rho A \frac{\partial^2 z}{\partial t^2}
    + EI\frac{\partial^4 z}{\partial x^4}
    - T\frac{\partial^2 z}{\partial x^2}
    = F(x,t),
\end{equation}
where the first term accounts for the inertia of the CNT, with $\rho$ the
mass density of the CNT and $A$ the cross-section area.
The second term accounts for the restoring force due to the bending rigidity
$EI$, with $E$ the Young modulus and $I$ the second moment of inertia,
while the third term is the restoring force due to the tension $T$.
Finally, the CNT is driven and tuned by the electrostatic force per unit length $F(x,t)$,
which is given by~\cite{witkamp2009thesis}
\begin{equation} \label{eq:fitm:Feq}
    F(x,t) = \frac{1}{2}\frac{\partial c_\mathrm{G}^l}{\partial z}(V_G(t) - V_0)^2,
\end{equation}
where $V_0$ is an offset on the dc gate voltage $V_G(t)$ and $c_\mathrm{G}^l = C_\mathrm{G}/L$, with $C_{\mathrm{G}}$ the total capacitance between the CNT and the gates, is the
gate capacitance per unit length.
If we approximate the geometry of the problem as that of a cylinder above
an infinite plane, then~\cite{witkamp2009thesis}
\begin{align} \label{eq:fitm:Ceq}
    c_\mathrm{G}^l(x) &= \frac{2\pi\epsilon_0}{\mathrm{arccosh}[(d - z(x))/r]}\nonumber\\
    &\approx 
    \frac{2\pi\epsilon_0}{\ln(2d/r)}
    +
    \frac{2\pi\epsilon_0}{\sqrt{d^2 - r^2}\mathrm{arccosh}^2(d/r)}z(x),
\end{align}
where $d$ is height of the CNT from the gates (at zero gate voltage),
$r$ is the radius of the CNT, and the last approximation is valid for
$d \gg r,z$.
Finally, the tension $T$ on the CNT has two contributions: one due to the
pull of CNT towards the gates which elongates it, and another due to clamping
which can introduce a residual tension $T_r$ and bending (so that the length of
the clamped CNT is not the same as the length when unclamped) even when the gate
voltage is zero. In conclusion, the tension is given by~\cite{witkamp2009thesis}
\begin{equation} \label{eq:fitm:Teq}
    T = T_r + \frac{E A}{2 L}\int_0^L\left(\frac{\partial z}{\partial x}\right)^2\mathrm{d}x.
\end{equation}

One can then get frequency of the eigenmodes of \eqref{eq:fitm:modeeq} by solving
\eqref{eq:fitm:modeeq} self-consistently together with \eqref{eq:fitm:Feq} and
\eqref{eq:fitm:Teq} (see~\cite{witkamp2009thesis} for details on these calculations).
We use the obtained fundamental frequency to fit the gate voltage dependence measured
in Fig.~\ref{fig:mass}.
To do the fit, we take $\rho = 1350\,\mathrm{kg}/\mathrm{m}^3$ and $E = 1.25\,\mathrm{TPa}$,
which are standard values for a CNT as it has been widely reported in the
literature~\cite{witkamp2009thesis,meerwaldt2012,castellanos-gomez2012}.
We further know from the device fabrication that $d \approx 100\,\mathrm{nm}$.
The parameters left to fit are then $L$, $r$, $T_r$, and $V_0$.
The obtained values are shown in Table~\ref{table:CalcMass} and the resulting
fit in Fig.~\ref{fig:mass} (dashed black line).
From $L$ and $r$ we further estimate the mass $m = 61 \pm 6\,\mathrm{ag}$.
This gives a zero point motion $\zpm = 0.68 \pm 0.04\,\mathrm{pm}$.
It is worth pointing out that the uncertainty on $\zpm$ does not affect the value of
coupling coefficient $\gm$ in the main text, since the expression of the effective
mechanical frequency [Eq.~\eqref{eq:softening}] does not depend on $\zpm$.
Furthermore, we have found that, in the numerical simulations described in the main
text, small changes in $\zpm$ only affect the value found for $z(0)$ which is such
that $z(0)/\zpm \simeq 20$.

\subsection{Estimation of the size of the quantum dot}\label{appendix:size}

Here we estimate the length $L_\mathrm{dot}$ of the quantum dot confinement in the carbon nanotube from the formula of the capacitance between a cylinder and an infinite plane, which for $d \gg z,r$ takes the form
\begin{equation}
    C_\mathrm{G}^\mathrm{dot} \approx \frac{2\pi\varepsilon_0L_\mathrm{dot}}{\ln{(2d/r)}}
\end{equation}
where $\varepsilon_0$ is the vacuum permittivity.
We reproduce the quantum dot capacitance with respect with $V_\mathrm{G1}$,  $C_\mathrm{G}^\mathrm{dot}=1.46$~aF, estimated from the Coulomb diamond in Fig.~\ref{fig:Experiments}(a), for $L_\mathrm{dot}=103$~nm and using the parameters of Table~\ref{table:CalcMass}.

\section{Full set of Coulomb peaks}\label{appendix:extra peaks}

\begin{figure*}[htb!]
    \centering
    \includegraphics[width=0.85\linewidth]{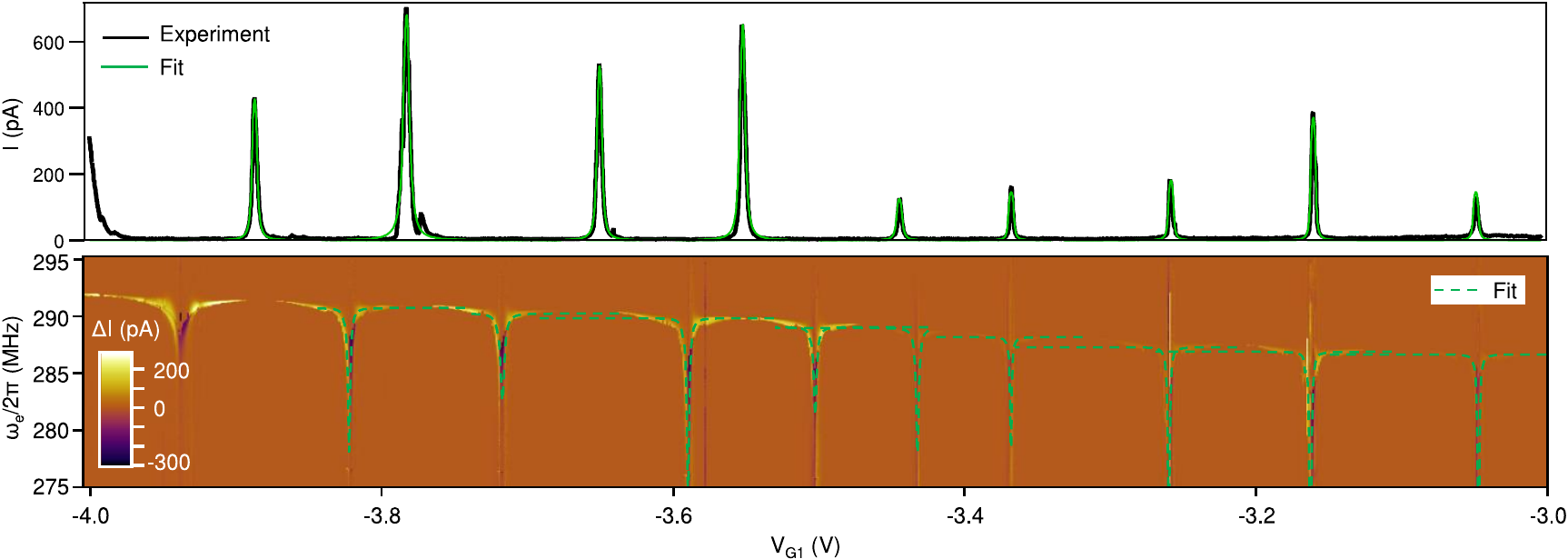}
    \caption{%
        Top: Fit (green line) of the Coulomb peaks measured under the same condition as in  Fig.~\ref{fig:Experiments}(b) for the extended range of $V_\mathrm{G1}$ (black line). The center of the Coulomb peaks in gate voltage are shifted from the centers of the frequency dips due to gate voltage drifts.
        Bottom: Corresponding mechanical resonance frequency obtained by measuring the current variation $\Delta I$. The plot was leveled by column for clarity. The mechanical resonance frequency variation are fitted using the linearization presented in Sec.~\ref{sec:model}, using $\gm$ as fitting parameter.
        The fitting parameters are displayed in Table \ref{table:OtherPeak}.}
    \label{fig:OtherPeaks}
\end{figure*}

We show in Fig.~\ref{fig:OtherPeaks}  other Coulomb peaks than the one studied in the main text. For each Coulomb peak, there is a dip in the mechanical resonance frequency. We applied the method described in the main text to estimate the coupling strength $\gm$. The results of each fit are summarized in Table~\ref{table:OtherPeak}. Note that in some cases the coupling strength we find exceeds the one in the main text, but the uncertainty is higher.
Our data shows no evidence of a dependence of $\gm$ with gate voltage. However, the range of gate voltage might be too small to reveal a trend.

\begin{table}[htb!]
    \begin{tabular}{ccccccc}
        \hline\hline
        Param. &        $\G_\mathrm{L}/2\pi$ &        $\G_\mathrm{R}/2\pi$ &        $V_\mathrm{G1}^\mathrm{f}$  &        $V_\mathrm{G1}^\mathrm{pk}$  &        $\omegam/2\pi$  &        $\gm/2\pi$ \\
        Unit & GHz & GHz & V &V & MHz& GHz \\
        \hline
        Peak 1 & 0.7 & 25 & -3.821 & -3.886 & 294.6 & 0.8\\
        Peak 2 & 1.7 & 40 & -3.716 & -3.7815 & 294.0 & 0.8 \\
        Peak 3 & 0.9 & 25 & -3.589 & -3.6485 & 293.5 & 0.9 \\
        Peak 4 & 1.3 & 30 & -3.502 & -3.55 & 292.5 & 0.7 \\
        Peak 5 & 0.2 & 30 & -3.432 & -3.442 & 292.4 & 0.8 \\
        Peak 6 & 0.2 & 20 & -3.368 & -3.365 & 291.4 & 0.8 \\
        Peak 7 & 0.25 & 20 & -3.260 & -3.255 & 290.3 & 0.8 \\
        Peak 8 & 0.5 & 15 & -3.163 & -3.157 & 289.8 & 0.8 \\
        Peak 9 & 0.2     & 20  & -3.055& -3.045 & 289.5      & 1.0  \\ 
        \hline\hline
    \end{tabular}
    \caption{\label{table:OtherPeak}
        Parameters of the fit of the measurements shown in Figure~\ref{fig:OtherPeaks} from Peak 1 (on the left) to Peak 9 (on the right). The centers of the Coulomb peaks $V_\mathrm{G1}^\mathrm{pk}$ are different in gate voltage from the centers of the frequency dips $V_\mathrm{G1}^\mathrm{f}$ due to gate voltage drifts.
        Note that a few coupling strengths exceed the one of the main text, but the uncertainty is higher.
    }
\end{table}

\section{Confirmation of the small amplitude limit}\label{appendix:numerical simualtions}

In Sec.~\ref{sec:model}, we did a first order expansion to obtain the effective mechanical frequency [Eq.~\eqref{eq:softening}] and used this expression to fit the experimental data [Fig.~\ref{fig:Experiments}(d)] and extract the value of $\gm$. Here, we go one step further and numerically integrate the equation of motion \eqref{eq:eq_of_motion} to confirm the value found for the coupling strength. This numerical integration requires to choose values for $\upmu_0,\, \gm$ and a set of initial conditions $(\dot{z}(0),\, z(0))$. We choose $\dot{z}(0) = 0$ and therefore $z(0)$ is closely related to the amplitude of the mechanical motion. The other parameters were determined from the experimental data and are given in Table \ref{table:params}.\\

\begin{figure}[h!]
    \includegraphics[width=0.95\linewidth]{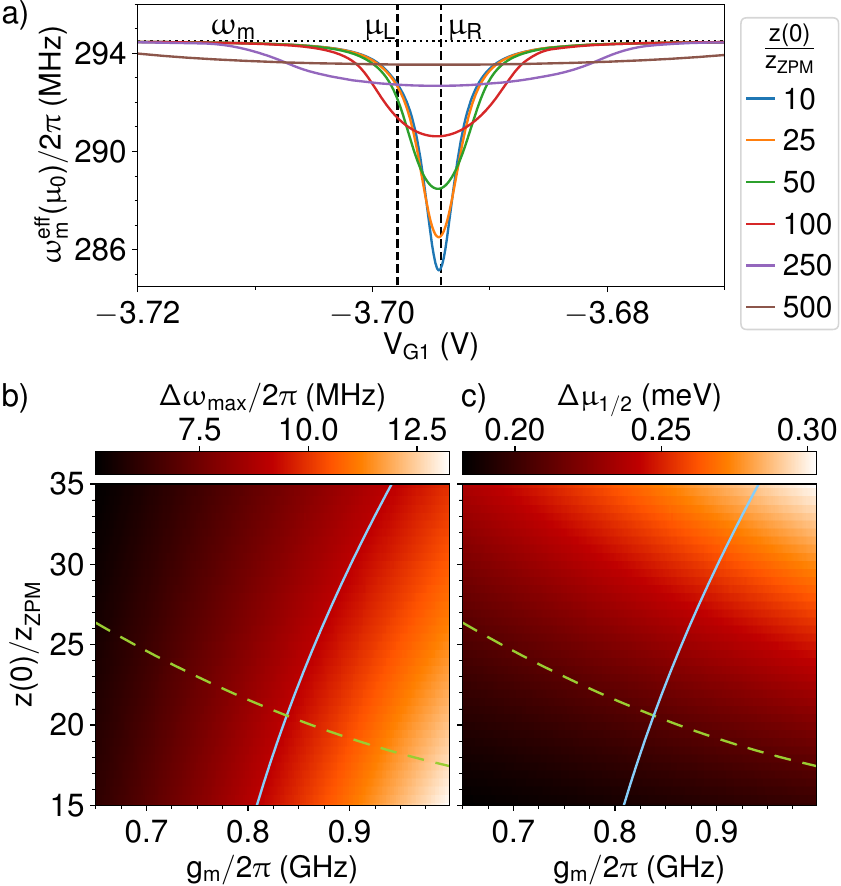}\\
    \caption{\label{fig:appendix_fit_simu}
        Characterization of the frequency dip: (a) mechanical resonance frequency as a function of the gate voltage $V_{G5} = V_0 - \upmu_0 / \alpha$ for $\gm/2\pi = 0.80$ GHz and different values of $z(0)$, (b) depth of the dip $\Delta\omega_\text{max} = \omegam - \min_{\upmu_0}(\omegam^\text{eff}(\upmu_0))$ and (c) full width at half-minimum $\Delta \upmu_{1/2}$ as a function of the initial position $z(0)$ and the coupling strength $\gm$. The light blue solid line corresponds to $\Delta\omega_\text{max} /2\pi = 9.1$ MHz and the dashed green line to $\Delta\upmu = 0.21$ meV, which are the characteristics of the experimental plot in Fig.~\ref{fig:Model}(c). The other parameters  are given in Table \ref{table:params}.
    }
\end{figure}

For each set of values $(\upmu_0, \gm, z(0))$, we get $\omegam^\text{eff}(\upmu_0)$ as the slope of the argument of $z(t) + i \dot{z}(t)/\omegam$.
Fig.~\ref{fig:appendix_fit_simu}(a) represents $\omegam^\text{eff}(\upmu_0)$ as a function of the gate voltage for $\gm/2\pi = 0.80$ GHz and different values of $z(0)$. The quantum dot level $\upmu_0$ is related to the gate voltage by the relation  $\upmu_0=\alpha\Delta V_\mathrm{G1})$,
k and $\alpha = 0.054$ eV/V. This figure shows that the frequency dip can be characterized by its depth, $\Delta\omega_\text{max} = \omegam - \min_{\upmu_0}(\omegam^\text{eff}(\upmu_0))$, and full width at half-minimum, $\Delta \upmu_{1/2}$. 
We extract these two parameters from the experimental data: $\Delta\omega_\text{max}^\text{exp}/2\pi = 9.1$ MHz and $\Delta\upmu_{1/2}^\text{exp} = 0.21$ meV. We then plot maps of $\Delta\omega_\text{max}$ and $\Delta \upmu_{1/2}$ [Fig.~\ref{fig:appendix_fit_simu} (b) and (c) respectively] as functions of $\gm$ and $z(0)$. The solid light-blue line corresponds to the experimental depth and the dashed yellow line to the width at half-minimum. The intersection of the two curves gives us the coupling strength: $\gm/2\pi = 0.84$ GHz for $z(0) = 21\zpm$. With this method, we obtain a coupling strength in good agreement with the analytical fit ($\gm = 0.80\pm0.04$ GHz), thus validating the first order expansion and, in addition we get an estimate of the amplitude of the mechanical oscillations, $\sim15$ pm.\\

In Fig.~\ref{fig:appendix_fit_simu}(a), we note that the dips are centered on the chemical potential of the right reservoir, $\muR$. This is because the two  barriers have very different tunnel rates: $\G_L \ll \G_R$. In addition, for small amplitudes of the mechanical oscillations, the widths of the dips are very similar, with $\Delta\upmu_{1/2} \simeq \hbar\Gb$. In this limit, the effective frequency is well estimated by Eq.~\eqref{eq:softening}. Conversely, the frequency dip becomes larger when the amplitude of the mechanical oscillations makes $\upmu(z)$ vary more than $\hbar\Gb$, that is for $z(0)/\zpm > \Gb/\gm$. In this case, $\upmu(z)$ can enter the bias window even for a $\upmu_0$ relatively far outside. Note that $\hbar\Gb$ is the length of the interval centered in $\muR$ over which $p(\upmu)$ varies significantly, see Eqs.~\eqref{eq:rho} and \eqref{equ:pop}. For the experimental device, we have $\Gb/\gm = 50$ so we can reasonably use the small amplitude limit.

\section{Electric field and single-particle energy levels}\label{appendix:electrostatics}

\begin{figure}[htb!]
    \centering
    \includegraphics[width=0.75\linewidth]{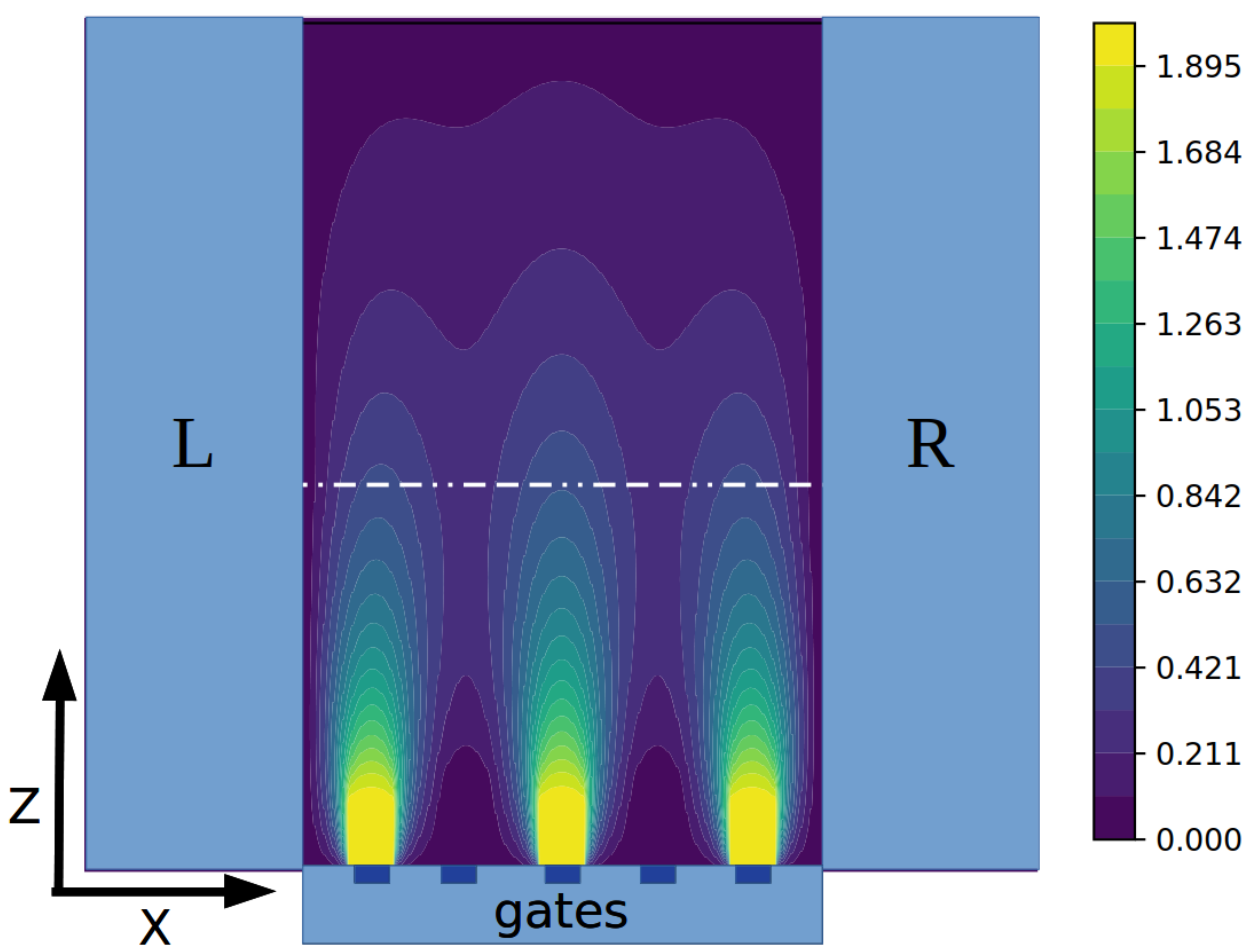}
    \caption{Electrical voltage in the vertical plane of the device in an arbitrary gates configuration. The nanotube is represented with an horizontal white line between the electrodes L and R, at a certain distance from the gates. This line represents the integration path of Eq.(11) of the main text.}
    \label{fig:electromagnetic2}
\end{figure}

We used a finite-differences method in order to calculate the electric potential field in a vertical plane on the device. In this calculation we considered a $1200\times1200$ grid in order to obtain enough resolution, and imposed the five gates in the bottom on the figure and the two lateral electrodes. The top boundary of the device is considered at sufficient height from the device and kept at constant zero voltage, obtaining a negligible impact on the system \cite{heinze2003unexpected, heinze2002carbon}.

Equation \eqref{eq:bohr} requires a path integral along the classical electrons trajectory. In Fig.~\ref{fig:electromagnetic2}, we sketch this situation. The electrons path representing the nanotube is considered as a straight horizontal line at a certain distance from the gates, moving from left lead to right.


\section{Estimation of the coupling strength for similar devices in the literature}\label{appendix:other devices}

\begin{figure}[htb!]
    \centering
    \includegraphics{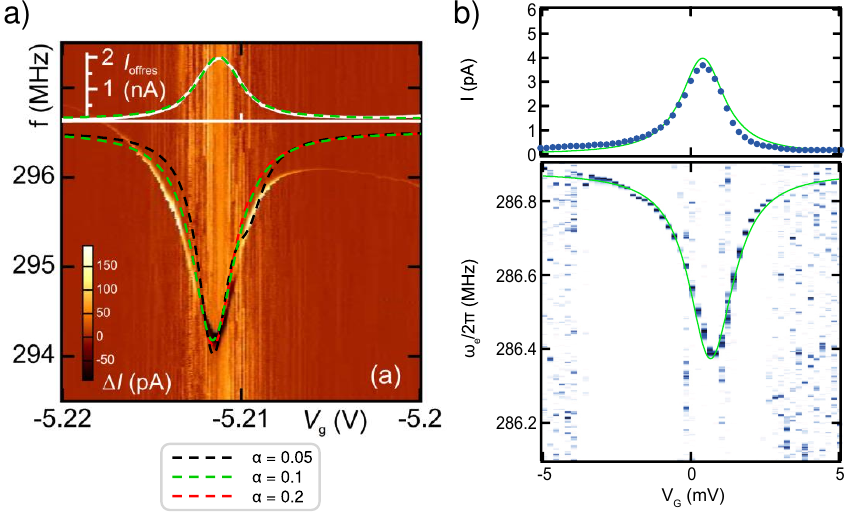}
    \caption{Fits of the Coulomb peak (upper plots) and mechanical resonance frequency (lower plots) for experimental results from other groups using our model.
        (a) Results reproduced from Fig.~4 of Ref.~\cite{Huttel2010}. The three curves correspond to different values of the lever arm: 
        $\alpha = 0.05$ ($\GL/2\pi=20$~GHz, $\GR/2\pi=5$~GHz, $\gm/2\pi = 0.34$~GHz),
        $\alpha = 0.1 $ ($\GL/2\pi=55$~GHz, $\GR/2\pi=10$~GHz, $\gm/2\pi = 0.5$~GHz) and 
        $\alpha = 0.2 $ ($\GL/2\pi=110$~GHz, $\GR/2\pi=19$~GHz, $\gm/2\pi = 0.7$~GHz).
        (b) Results reproduced from Fig.~4 of Ref.~\cite{Steele2012}. The parameters of the fit were $\G_\mathrm{L}/2\pi = 127$~GHz, $\G_\mathrm{R}/2\pi = 30$~GHz, $\gm/2\pi = 0.36$~MHz. The mechanical resonance frequency $\omegam/2\pi = 286.88$~MHz and the lever arm $\alpha=0.38$ are given in the paper. The Coulomb peak is shifted by 0.8~mV between the top and bottom panel.
    }
    \label{fig:Huttel and Steele}
\end{figure}

We apply our model to fit measurements obtained with similar devices \cite{Huttel2010,Steele2012} and show that devices in these studies are also in the ultrastrong coupling regime. First we fit the results of A.K. Huttel \etal~\cite{Huttel2010} in Fig.~\ref{fig:Huttel and Steele}(a). We find $\gm/2\pi \approx 0.5$~GHz. The other parameters of the fit are $\GL=55$~GHz, $\GR=10$~GHz and $\omegam/2\pi=296.5$~MHz, considering $\Vs=0.1$~mV from the paper and $\alpha=0.1$. The lever $\alpha$ is not mentioned so we explored a range of value from $\alpha=0.05$ to $\alpha=0.2$. The best fit is obtained with $\alpha=0.1$.
We also fitted the result published by Meerwaldt \etal~\cite{Steele2012} that display all the parameters our model requires, see Fig.~\ref{fig:Huttel and Steele}(b). We find a coupling of $\gm/2\pi = 0.36$~GHz for $\G_\mathrm{L}/2\pi = 127$~GHz, $\G_\mathrm{R}/2\pi = 30$~GHz and $\omegam/2\pi = 286.88$~MHz.

In conclusion we find a coupling strength of $\gm/2\pi \approx 0.5$~GHz and $\G_\mathrm{R}/2\pi = 30$~GHz in these two devices giving a $\gm/\omegam$ ratio of 1.7 and 1.25, respectively.

\bibliography{main.bib}

\end{document}